\documentclass[reprint,twocolumn, amsmath,amssymb,amsfonts, aps, prl, superscriptaddress,notitlepage,longbibliography]{revtex4-2}

\usepackage{graphicx}
\usepackage{physics}
\usepackage{float} 
\usepackage{subfigure} 
\usepackage{color}
\usepackage{amsmath}
\usepackage{amsfonts}
\usepackage[normalem]{ulem}
\usepackage{xspace}
\usepackage[dvipsnames]{xcolor}
\usepackage{dcolumn}
\usepackage{setspace}
\usepackage[breaklinks,colorlinks,linkcolor=blue,citecolor=blue,urlcolor=blue]{hyperref}

\begin{document}
\title{Polarization-dependent chiral transport and chiral solitons in spin Kitaev models}

 \author{Chenwei Lv}
 \affiliation{Department of Physics and Astronomy, Purdue University, West Lafayette, IN, 47907, USA}
\affiliation{Department of Physics, The Chinese University of Hong Kong, Shatin, New Territories, Hong Kong, China}

 \author{Thomas Bilitewski}
\affiliation{Department of Physics, Oklahoma State University, Stillwater, Oklahoma 74078, USA}
 
\author{Ana Maria Rey}
\email{arey@jila.colorado.edu}
 \affiliation{JILA, NIST and Department of Physics, University of Colorado, Boulder, CO, USA}
 \affiliation{Center for Theory of Quantum Matter, University of Colorado, Boulder, CO, USA}

\author{Qi Zhou}
\email{zhou753@purdue.edu}
\affiliation{Department of Physics and Astronomy, Purdue University, West Lafayette, IN, 47907, USA}
\affiliation{Purdue Quantum Science and Engineering Institute, Purdue University, West Lafayette, IN 47907, USA}

\date{\today}

\begin{abstract}
Recent advances in synthetic quantum matter allow researchers to design quantum models inaccessible in traditional materials. Here, we propose protocols to engineer a new class of quantum spin models, which we call spin Kitaev models. The building blocks are basic spin-exchange interactions combined with locally selective Floquet pulses, a capability recently demonstrated in a range of experimental platforms. The resulting flip-flip and flop-flop terms lead to intriguing quantum transport dynamics beyond conventional spin models. For instance, in the absence of a magnetic field, spin excitations polarized along the $x$ and $y$ axes propagate chirally in opposite directions, producing polarization-dependent spin transport. In the large-spin limit, the spin Kitaev model maps to a nonlinear Hatano-Nelson model, where the interplay of nonlinearity and the underlying curvature yields polarization-dependent chiral solitons. A magnetic field binds two oppositely polarized chiral solitons into a chiral solitonic molecule, whose travel direction depends on its orientation. Our results, directly accessible in current experiments, open new opportunities for simulating transport in curved spaces and for applications in spintronics, information processing, and quantum sensing.
\end{abstract}
\maketitle 

\begin{spacing}{0.98}
Over the past decade, synthetic quantum matter has emerged as a powerful frontier in quantum science, enabling controlled exploration of many-body phenomena and quantum phases inaccessible in natural materials. 
A central driver of this progress has been the rapid development of ultracold atomic and molecular platforms. 
Experimental advances include the preparation of quantum degenerate gases of magnetic atoms \cite{Chomaz2023}, association of ground-state polar molecules~\cite{Ni2008, Zwierlein2015}, the production of both molecular degenerate Fermi gases~\cite{DeMarco2019, Duda2023, Bigagli2024, Langen2024, Will2023} and Bose Einstein Condensates \cite{Bigagli2024Observation}, and the manipulation of the rich internal states of polar molecules~\cite{Rey2013, Yan2020, Matsuda2020, Anderegg2021, Schindewolf2022, Rey2023, Cornish2024} and magnetic atoms~\cite{Chomaz2023} for engineering tunable spin exchange interactions via electromagnetic fields~\cite{Bohn2017, Moses2017, Chomaz2023} or microwave control. 
In parallel, optical tweezers~\cite{Kaufman2021,anderegg2019optical, Holland2023} and Rydberg atoms provide complementary control of spin-spin interactions with single-particle resolution~\cite{Browaeys2020, Scholl2022, Nishad2023, Saffman2010}. 
Floquet engineering offers an additional layer of controllability~\cite{Dalibard2014, Bukov_2015, Oka_2019, Eckardt2017, Choi2020}, and has recently been demonstrated in both molecular~\cite{Christakis2023, Miller2024} and Rydberg platforms~\cite{Geier2021, Scholl2022}, enabling the realization of diverse spin Hamiltonians, including Ising, anisotropic Heisenberg, and generalized XYZ models~\cite{Lukin_2020_Robust, Geier2021, Scholl2022, Christakis2023, Miller2024}. 
Altogether, the developments set the stage for engineering new classes of spin models that go beyond conventional paradigms.
\end{spacing}

\begin{figure}[ht]
    \centering
    \includegraphics[width=0.45\textwidth]{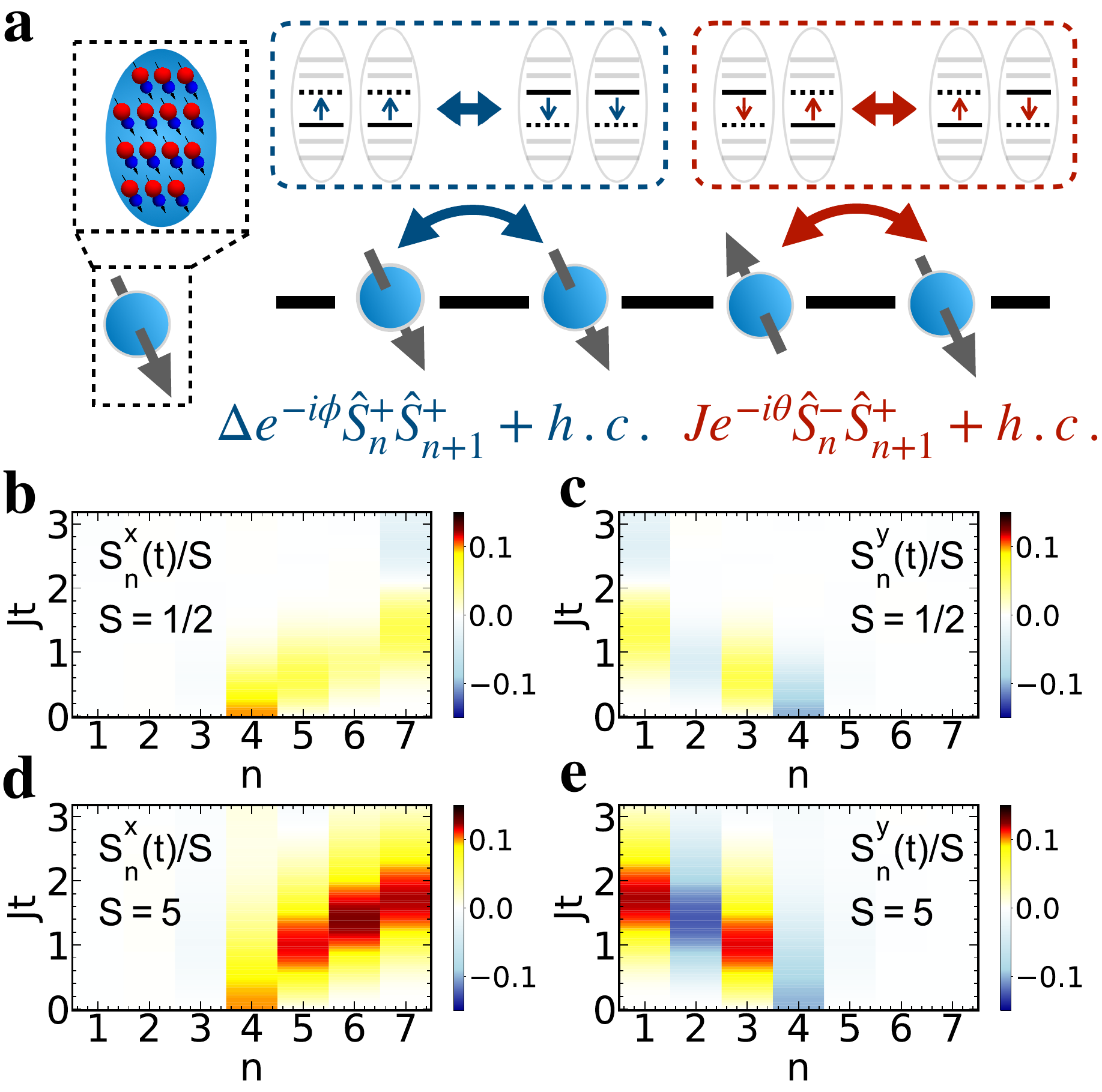}
    \caption{
    (a) Schematic plot illustrating how to realize a large spin from an ensemble of many spin 1/2 particles (left) and the interaction terms of the spin Kitaev model (right). Polar molecules are shown as an example, similar schemes also apply to magnetic and Rydberg atoms.
    (b-e) Polarization-dependent chiral spin transport. $S^x_{n}(t)/S$ (b,d) and  $S^y_{n}(t)/S$ (c,e) as functions of time for spin $S=1/2$ (b,c) and $5$ (d,e). Initial excitations created by $\hat U_{x,y}$ on $|S\rangle^N$ are located at $n_0=4$. $\Delta=0.8J$ and $N=7$ have been used. 
    \label{Fig:dynamics}
    }
\end{figure}

Motivated by these impressive experimental developments, here we propose and study a new class of one-dimensional quantum spin-$S$ models, which are written as $\hat H_{\rm sK}=\hat H_{\mathrm{ex}}+\hat H_{\rm ff}$,
\begin{equation}
\begin{split}
    \hat H_{\mathrm{ex}}&=\sum\nolimits_n(Je^{-i\theta}\hat{S}^-_n\hat{S}^+_{n+1}+h.c.)/(2S),\\
     \hat H_{\rm ff}&= \sum\nolimits_n(\Delta e^{-i\phi} \hat{S}^+_n\hat{S}^+_{n+1}+h.c.)/(2S),
     \label{eq:H_sK}
\end{split}
\end{equation}
schematically illustrated in Fig.~\ref{Fig:dynamics}(a). 
Here, $(\hat{S}^x_n,\hat{S}^y_n, \hat{S}^z_n)$ is the spin angular momentum operator for a spin-$S$ at site $n$ and $\hat{S}^{\pm}_n=\hat{S}^x_n\pm i\hat{S}^y_n$. 
Similar to conventional quantum spin models, $\hat H_{\rm sK}$ includes a spin-exchange (or flip-flop) interaction, $Je^{-i\theta}\hat{S}^-_n\hat{S}^+_{n+1}$ and its Hermitian conjugate, where $J$ and $\theta$ are the amplitude and its phase, respectively. 
In addition, $\hat H_{\rm sK}$ includes the flip-flip term $\Delta e^{-i\phi}\hat{S}^+_n\hat{S}^+_{n+1}$ and the flop-flop term $\Delta e^{+i\phi}\hat{S}_n^-\hat{S}_{n+1}^-$, with an amplitude $\Delta$ and a phase $\phi$, respectively.
Most common experimental realizations conserve the total axial magnetization, $S_z=\sum_n S^z_n=\sum_n\langle \hat{S}^z_n\rangle$, and therefore the flip-flip and flop-flop terms are forbidden, and not easily accessible. 
Some exceptions are implementations using motional levels as an effective pseudo-spin degree of freedom. 
For example, using $s$ and $p$ bands in a double-well lattice, the interaction-induced hybridization can generate terms, $(\hat{S}^+_n\hat{S}^+_{n+1}+h.c.)$ with  $\theta=\phi=0$~\cite{Zhou2011}, or multiple tones in momentum states in an optical cavity\cite{Luo2025Hamiltonian}. 

In this work, we present an experimentally accessible way to realize $\hat H_{\rm sK}$ for arbitrary spin quantum number $S$ and arbitrary phases $\theta$ and $\phi$ using Floquet-engineered interactions between spin 1/2 particles. 
Our scheme is accessible in state-of-the-art experimental platforms, including magnetic atoms, molecules, and Rydberg atoms in optical lattices and optical tweezers under Floquet driving. 
Building on this, we uncover novel chiral transport behaviors governed by $\hat H_{\rm sK}$.  
Previous studies explored chiral transport in the Hatano-Nelson model (HNM)~\cite{Hatano1996, Hatano1998, Weidemann2020, Yan2022}, a non-Hermitian tight-binding Hamiltonian written in terms of bosonic or fermionic operator, where asymmetric tunneling leads to directional amplification, and in the Bosonic Kitaev model (BKM)~\cite{Clerk2018, Clerk2020}, where quadrature-dependent chiral transport arises from its decomposition into two Hatano-Nelson chains. 
Our work reveals a distinct mechanism: the interplay between exchange and flip-flip/flop-flop interactions in $\hat H_{\rm sK}$ gives rise to polarization-dependent chiral transport for arbitrary $S$. 

We find that in a fully polarized spin Kitaev chain, local excitations in the $x$ and $y$ spin projections amplify in opposite directions. 
Furthermore, the intrinsic nonlinearity of $\hat H_{\rm sK}$ leads to intriguing phenomena absent in both HNM and BKM.
In our setting, the density-dependent mass leads to a nonlinear HNM and the emergence of polarization-dependent chiral solitons and solitonic molecules that propagate unidirectionally without dispersion. 
These findings open new avenues for exploring chiral transport and nonlinear dynamics in quantum systems, with experimental feasibility in current platforms and potential applications in quantum spintronics, quantum networks, and robust quantum interferometry~\cite{Xiao2015, Chumak2015, Yang2021, Zoller2015, Polo2013, Robins2014}.

{\it Floquet Implementation---} 
To engineer the spin Kitaev model $\hat H_{\rm sK}$ in Eq.~(\ref{eq:H_sK}) with arbitrary $\theta$ and $\phi$, we start from the spin-exchange Hamiltonian $\hat{H}'_{\rm{ex}}=(J_0/2S)\sum_n(\hat{S}^+_n\hat{S}^-_{n+1}+h.c.)$, naturally realized in a wide range of experimental platforms. 
We then apply Floquet engineering techniques to $\hat{H}'_{\rm ex}$ ~\cite{Dalibard2014,Bukov_2015, Eckardt2017, Oka_2019, Choi2020} in the spirit of experimentally demonstrated protocols realizing the XYZ model~\cite{Geier2021, Scholl2022, Nishad2023, Miller2024}. 

By using locally controlled Stark-shifts, or staggered electric fields,  $\hat{\mathcal{U}} = e^{-i\sum_{m}m \hat S_m^z\pi/2}$, $\hat{\mathcal{W}}=\prod_{m\in 2\mathbb{Z}}e^{-i\hat S_m^z \pi/2}$,  and site-selective rotations acting on even sites implementable through Raman transitions, $\hat{\mathcal{V}}=\prod_{m\in 2\mathbb{Z}}e^{-i\hat S_m^x \pi}$, we find the following pulse sequence 
\begin{equation} \label{Eq:Sequence}
  \begin{split}
        \hat U &= \hat{\mathcal{U}} e^{-i\tilde  H_{\mathrm{ex}}\tau_4}\hat{\mathcal{U}}^\dag e^{-i\tilde  H_{\mathrm{ex}}\tau_1}\hat{\mathcal{V}}^\dag e^{-i\tilde H_{\mathrm{ex}}\tau_2}\hat{\mathcal{W}}^\dag \\ 
               &\qquad e^{-2i\tilde H_{\mathrm{ex}}\tau_3}\hat{\mathcal{W}} e^{-i\tilde  H_{\mathrm{ex}}\tau_2}\hat{\mathcal{V}} e^{-i\tilde H_{\mathrm{ex}}\tau_1}\hat{\mathcal{U}}e^{-i\tilde  H_{\mathrm{ex}}\tau_4}\hat{\mathcal{U}}^\dag,
  \end{split}
\end{equation}
delivers an effective $\hat H_{\rm sK}$ with $J e^{-i\theta}=2J_0(\tau_1-i\tau_4)/T$ and $\Delta e^{-i\phi}=2J_0(\tau_2-i\tau_3)/T$, where $T=2\sum_i\tau_i$ is the driving period (Supplementary Materials~\cite{supplemental}).

While this scheme applies to any generic spin-$S$, most experimental platforms do not naturally host large spin-$S$ particles. 
Therefore, to simulate $\hat{H}_{\text{sK}}$ for any $S$, we consider each site to correspond to a two-dimensional plane or one-dimensional chain hosting multiple spin-$1/2$'s, as shown schematically in Fig.~\ref{Fig:dynamics}(a). 
In the presence of strong in-plane Heisenberg interactions, $N$ spin-$1/2$'s can behave like a collective spin with $S=N/2$~\cite{Rey2008, Martin2013, Bilitewski2023} during the quantum dynamics of initially fully polarized states. 
We extend the described Floquet protocol to engineer both the in-plane Heisenberg interactions and out-of-plane interactions~\cite{supplemental}, thus enabling experimental access to the physics of an arbitrary spin-$S$ Kitaev model starting from easily accessible spin-exchange interactions between spin-$1/2$'s arranged in multi-layers.

{\it Chiral Transport---}
We now turn to transport phenomena of $\hat H_{\rm sK}$ focusing on $\theta=\phi=\pi/2$, which is sufficient to illustrate the full phenomenology, and provide results for other $\theta$ and $\phi$ in the Supplementary Materials~\cite{supplemental}. 
We consider a fully polarized state $|F\rangle=\prod_n|S\rangle_n$ and implement a local excitation via $\hat{U}_x=e^{-i\alpha \hat{S}^y_{n_0}}$ or $\hat{U}_y=e^{-i\alpha \hat{S}^x_{n_0}}$ at site $n_0$ on $|F\rangle$, resulting in propagation and dynamics. 
$\hat{U}_x$ ($\hat{U}_y$) rotates the $n_0$-th spin about the $y$ ($x$) axis by $\alpha$ and thus creates a local excitation of $S^x_{n_0}$ ($S^y_{n_0}$), where $S^{x,y}_{n} = \langle \hat S^{x,y}_{n}\rangle$. 
We note that excitations in $S^x_{n}$ and $S^y_{n}$ are decoupled, after $\hat{U}_x$ ($\hat{U}_y$) creates a finite $S_x$ ($S_y$) from $|F\rangle$ at site $n_0$,  $S_y$ ($S_x$) remains zero. 

For finite $S$, we numerically compute the dynamics and evaluate $S^x_{n}(t)$ and  $S^y_{n}(t)$ as functions of time $t$, for a range of models with spin $S=1/2$ up to $S=5$ in a lattice of $N=7$ sites using exact diagonalization (Supplementary materials~\cite{supplemental}). 
The results for the time evolution of these observables for $S=1/2$ and $S=5$ are shown in Fig.~\ref{Fig:dynamics}(b-e), where $\theta=\phi=\pi/2$, $\Delta=0.8J$ is chosen such that the asymmetric transport is evident even for small $N=7$, and we use $n_0=4$ and $\alpha = 0.1$. 
For both values of $S$, we clearly observe asymmetric spin transport: $S_x$ excitations dominantly propagate to the right (b and d), and $S_y$ excitations propagate to the left (c and e).
During time evolution, the spin excitations get amplified toward a unique direction, a characteristic feature of the skin effect~\cite{Weidemann2020, Yan2022}. 
Since the amplification of $S_x$ and $S_y$ excitations occur in opposite directions, we conclude that $\hat H_{\rm sK}$ exhibits a polarization-dependent skin effect. 

The numerics become significantly more challenging for larger $S$. 
However, we can gain analytical insight in the limit $S\rightarrow \infty$. 
Applying the Holstein-Primakoff transformation to leading order, $\hat S_n^+ = \sqrt{2S}\hat a_n$, $\quad \hat S_n^- = \sqrt{2S}\hat a^\dag_n$, $\hat H_{\rm sK}$ reduces to the BKM,
\begin{equation}
\hat H_{\rm bK}= \sum_{n} Je^{-i\theta} \hat a_n^\dag \hat a_{n+1} + \Delta e^{-i\phi}\hat a_n \hat a_{n+1} +  h.c.~\label{Eq:BKC}.
\end{equation}
Focusing again for simplicity on the parameters $\theta=\phi=\pi/2$ and defining the quadratures, $\hat{S}^x_n\rightarrow \hat X_n=(\hat{a}^\dagger_n+\hat{a}_n)/\sqrt{2}$ and $\hat{S}^y_n\rightarrow \hat P=i(\hat{a}^\dagger_n-\hat{a}_n)/\sqrt{2}$, and their expectation values $X_n = \langle \hat X_n\rangle$ ($P_n = \langle \hat P_n\rangle$), we obtain 
\begin{equation}
\begin{split}
    &\partial_t X_n=(\Delta+J)X_{n-1}+(\Delta-J) X_{n+1},\\
    &\partial_t P_n=(J-\Delta)P_{n-1}-(\Delta+J) P_{n+1},\label{eq:HN_quadratures}
\end{split}
\end{equation} 
each of which corresponds to a Hatano-Nelson chain with nonreciprocal couplings with $\Delta/J$ controlling the asymmetry of the couplings~\cite{Clerk2018}. 

Excitations in the $X$-quadrature and $P$-quadrature propagate and get amplified toward opposite directions, which is most easily seen when $\Delta = J$, i.e., the couplings in Eq.~(\ref{eq:HN_quadratures}) become unidirectional. 
More generally, the directional transport can be shown to originate from the finite curvature of an underlying hyperbolic space~\cite{lv2022curving, Zhou2024}. 
Our derivation of Eq.~(\ref{eq:HN_quadratures}) in the infinite $S$ limit therefore demonstrates that the quadrature-dependent chiral skin effect of the BKM belongs to a larger class of polarization-dependent chiral skin effects. 

When $S$ is finite, the spin operator cannot be written in terms that are linear in $X_n$ and $P_n$. 
Instead, we can understand the polarization-dependent chiral skin effect to arise from interference between multiple pathways created by $\hat H_{\mathrm{ex}}$ and $\hat H_{\rm ff}$. This is most easily demonstrated in the early-time dynamics. 
For $\ket{\psi(0)}=\hat{U}_x|F\rangle$ as the initial state, it is sufficient to consider three sites, $n_0-1$, $n_0$, and $n_0+1$ at early times, as spins in all other sites still point upward. 
For a small $\alpha$, the initial state can be approximately written as 
\begin{equation}
    \ket{\psi(0)} = \ket{...}\otimes(\ket{S,S,S}+c_1\ket{S,S-1,S}),
\end{equation}
where $c_1=\alpha\sqrt{S/2}$, $\ket{S, S-1, S}$ is a short-hand notation for spin states in sites $n_0$ and $n_0\pm1$, $\ket{S, S-1, S}\equiv\ket{S}_{n_0-1}\ket{S-1}_{n_0}\ket{S}_{n_0+1}$, $\ket{...}\equiv\prod_{m\neq n_0,n_0\pm 1}|S\rangle_m$. 

At small $t$ to first order $e^{-i\hat H_{\rm sK}t}\approx 1-i (\hat H_{\mathrm{ex}}+\hat H_{\rm ff})t$. 
We can interpret the action of $\hat H_{\mathrm{ex}}$ and $\hat H_{\rm ff}$ as each creating a pathway for spin transport. 
The time-evolved state reads $ \ket{\psi(t)} = \ket{...}\otimes\left(\psi_0+\psi_1(t)+\psi_2(t)\right)$, where 
\begin{equation}
    \begin{split}
        &\psi_0 = \ket{S,S,S}+c_1\ket{S,S-1,S},\\
        &\psi_1(t)= Jt c_1( \ket{S,S,S-1}-\ket{S-1,S,S} ),\\
        &\psi_2(t)=\Delta t(\ket{S-1,S-1,S}+\ket{S,S-1,S-1}).
    \end{split}
\end{equation}
Both $\psi_1(t)$ and $\psi_2(t)$ contribute to $S^{x,y}_n$. 
For instance, ${S}^{x}_{n_0\pm 1}=S\alpha t(\Delta \pm J)$. 
We see that a finite $J$ and a finite $\Delta$ lead to a constructive (destructive) interference at site $n_0+1$ ($n_0-1$). 
As such, ${S}^{x}_{n}$ gets amplified during propagation towards the right. Similarly, if we consider $\ket{\psi(0)}=\hat{U}_y|F\rangle$ as the initial state, ${S}^{y}_{n_0\pm 1}=S\alpha t(\Delta \mp J)$, which means that the interference becomes constructive (destructive) at site $n_0-1$ ($n_0+1$) and ${S}^{y}_{n}$ gets amplified during propagation towards the left.

Since this simple argument only applies at short times $Jt\ll 1$, and interference between more complex states and pathways is required at later times, it is difficult to obtain simple analytical results for the full time dynamics. 
However, our numerical results shown in Fig.~\ref{Fig:dynamics}(b-e), clearly demonstrate that the polarization-dependent chiral spin transport indeed persists beyond the short-time limit. 
Fig.~\ref{Fig:dynamics} also shows that the polarization-dependent chiral spin transport holds for generic $S$. 

Unlike for $\hat H_{\rm bK}$, which corresponds to infinite $S$, in the spin Kitaev model, the growth of a spin excitation is fundamentally bounded by the finite size of the spins $S$. 
Retaining nonlinearities is crucial to capture the correct dynamics of the finite $S$ model. 
We therefore explore the role of nonlinear terms in $\hat H_{\rm sK}$. 
We consider a large but finite $S$, where the semiclassical approach applies. 
The equations of motion are written as 
\begin{equation}  \label{Eq:semiS}
\begin{split}
    \partial_t J_{n}^{x} =& (\Delta + J)J_{n-1}^{x}J_{n}^{z} +(\Delta - J)J_{n+1}^{x}J_{n}^{z},\\
    \partial_t J_{n}^{y} =& (J - \Delta)J_{n-1}^{y}J_{n}^{z} -(\Delta + J)J_{n+1}^{y}J_{n}^{z},
\end{split}
\end{equation}
where $J^{x(y,z)}_{n}=S^{x(y,z)}_{n}/S$.
When only $S_n^x$ is excited,  $J_n^y$ remains zero, $J_{n}^{z}$ in the first line of Eq.~(\ref{Eq:semiS}) can be written as $J_{n}^{z}=\sqrt{1-(J_{n}^{x})^2}$ and we obtain
\begin{equation}
    \partial_t J_{n}^{x} =(1-(J_{n}^{x})^2)^{\frac{1}{2}}[(\Delta + J)J_{n-1}^{x} +(\Delta - J)J_{n+1}^{x}].
    \label{Eq:nonlinearHN}
\end{equation}
If the nonlinear term $(1-(J_{n}^{x})^2)^{\frac{1}{2}}$ (or $J_{n}^{z}$ in Eq.~(\ref{Eq:semiS})) is replaced by 1, the above equation reduces to the standard HNM as we obtained in Eq.~(\ref{eq:HN_quadratures}). 
As such, Eq.~(\ref{Eq:nonlinearHN}) is seen to be a nonlinear generalization of the HNM. The nonlinear term $(1-(J_{n}^{x})^2)^{\frac{1}{2}}$ can be interpreted as a density-dependent mass for the excitation.

\begin{figure}[th]
    \includegraphics[width=1\linewidth]{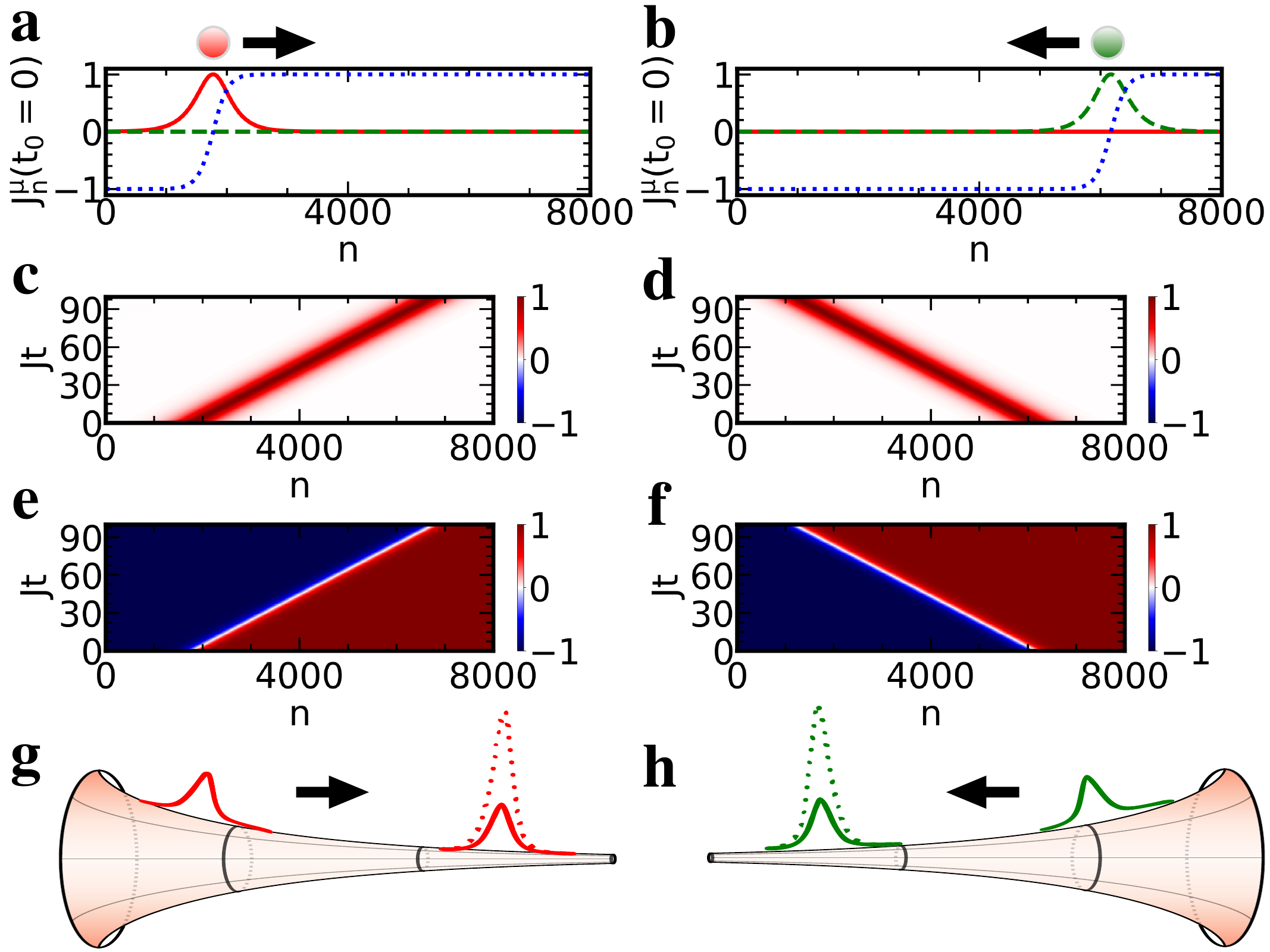}
    \caption
    {
    (a,b) Initial profiles of the (a) $x$-soliton (b) $y$-soliton. Red solid, green dashed, and blue dotted curves are $J^x_n(t)$, $J^y_n(t)$, and  $J^z_n(t)$ at $t_0=0$, respectively. 
    (c,e) Density plots of $J^x_n(t)$ and $J^z_n(t)$ of an $x$-soliton propagating towards right with $v_s=+50J$. $\Delta = 0.1J$.
    $J^y_n(t) = 0$.
    (d,f) Density plots of $J^y_n(t)$ and $J^z_n(t)$ of a $y$-soliton propagating towards left with $v_s=-50J$. $\Delta = 0.1J$.
    $J^x_n(t) = 0$.
    (g,h) are schematic plots of the $x$ and $y$-solitons in the curved spaces, where dotted curves denote the solitons in the absence of nonlinearity. 
    \label{fig:soliton}
    }
\end{figure}

{\it Chiral Solitons---}
We next explore the consequences of the nonlinear nature of the equations derived above, which we show result in the existence of chirally propagating solitonic excitations, i.e. wave-packets of spin excitations that propagate uni-directionally without dispersion in their shape as seen in Fig.~\ref{fig:soliton}.

Specifically, Eq.~(\ref{Eq:nonlinearHN}) admits soliton solutions that satisfy $J^\mu_n(t)=J^\mu_{n-v_st_0}(t-t_0)$, where $\mu=x,y,z$ and $v_s$ is the velocity of the soliton.  
We show the initial profiles, $J_n^{\mu=x,y,z}(t=t_0)$, of solitonic solutions in Fig.~\ref{fig:soliton} (a,b), and provide details of their construction in the Supplementary Materials~\cite{supplemental}.  
As observable in the figure, the solitons are well-localized wave-packets of spin-excitations, i.e. the profile $J_{n}^{x,y}$ is significant only within a finite window of lattice sites, while as $n\rightarrow\pm\infty$, $J_{n}^{z}\rightarrow \pm 1$, and $J_{n}^{x,y}$ vanishes. 
Far away from the maximum, $J_{n}^{x,y}$ is small, the nonlinearity is negligible, and the result recovers that of the conventional HNM~\cite{supplemental}.

Starting from these soliton profiles, we next study the time evolution of $J_n^\mu(t)$, by numerically solving Eq.~(\ref{Eq:nonlinearHN}). 
As shown in Fig.~\ref{fig:soliton}(c,e) for $x$ [and Fig.~\ref{fig:soliton}(d,f) for $y$-solitons], the soliton wave-packet $J_{n}^{x}$ ($J_{n}^{y}$) propagates towards the right (left) at constant velocity $v_s$ without dispersion as $t$ increases, a unique feature of chiral solitons. 
We find that these solitonic solutions only exist for $v_s>2J$, which guarantees that both the boundary conditions at $n\rightarrow \pm \infty$ are satisfied (Supplementary Materials \cite{supplemental}). 
With decreasing $v_s$, the width of the $x$-soliton decreases. 
When $v_s$ approaches $2J$, the spatial profiles of $J_n^\mu$ eventually become infinitely sharp and the solitonic solution disappears as $v_s \leq 2J$ (Supplementary Materials~\cite{supplemental}). 

The chiral $x$- and $y$-solitons can be understood as arising from the hyperbolic space geometry embedded in the HNM, and emerging as a consequence of the asymmetric coupling~\cite{lv2022curving}. 
In this case, as shown in Fig~\ref{fig:soliton}(g,h), when a wave-packet travels towards the direction with smaller circumferences, its amplitude increases to conserve the total probability~\cite{lv2022curving}.  
On the other hand, the polarization-dependent chiral soliton emerges due to the interplay between the hyperbolic space geometry and the nonlinearity.   
The latter, manifest as a density-dependent mass $(1-(J^x_n)^2)^{\frac{1}{2}}$ that stops the wave-packet growth and stabilizes solitons as depicted by the solid curves in Fig.~\ref{fig:soliton}(g,h).  

Note that the underlying mechanism of the chiral solitons is distinct from that in spin-orbit coupled condensate, where the locking between spin and momentum leads to a momentum-dependent interaction~\cite{Nishino1998, Edmonds2013, Frlian2022}.  
Here, when $\Delta=0$, the curvature becomes zero and the chiral transport vanishes~\cite{lv2022curving}. 
On the other hand, without the nonlinear term, the wave-packet would amplify exponentially without an upper bound as shown by the dotted lines in Fig.~\ref{fig:soliton}(g,h). 
 
We emphasize that the travel directions of both $x$- and $y$-solitons depend on the boundary conditions at $\pm\infty$. 
Flipping the sign of $J_{n}^{z}$ such that $J_{n}^{z}\rightarrow \infty$ ($-\infty$) when $n\rightarrow -\infty$ ($\infty$), the $x$- and $y$-solitons travel towards left and right, respectively. 
This can be understood from Eq.~(\ref{Eq:semiS}). 
Changing the sign of $J_{n}^{z}$ amounts to adding an extra minus sign to $\partial_t J_{n}^{x,y}$ and Eq.~(\ref{Eq:semiS}) becomes its time-reversal. 
Both $x$- and $y$-solitons thus travel towards opposite directions (Supplementary materials~\cite{supplemental}). 

\begin{figure}[t]
    \includegraphics[width=1\linewidth]{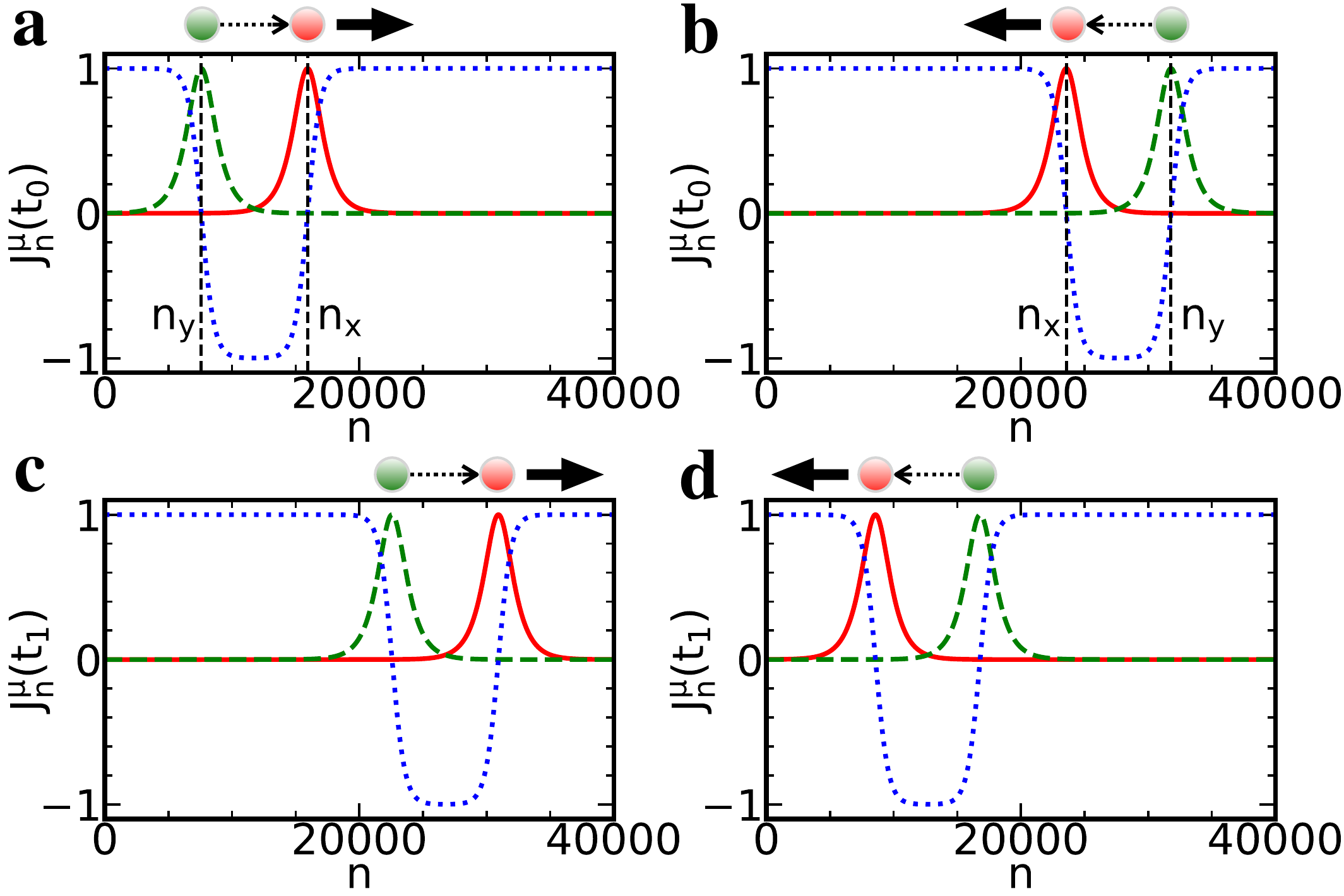}
    \caption{
    Solitonic molecule propagating towards left (a,c) and right (b,d) with $v_s=\pm 200J$. $\Delta = 0.1J$, and $h=-0.0001J$ are used. 
     Solid red, green dashed and blue dotted curves represent $J^x_n(t)$, $J^y_n(t)$ and $J^z_n(t)$ when $Jt_0=0$ (a,b) and $Jt_1=75$ (c,d). 
     Small dotted arrows denote the polarizations of solitonic molecules.
     \label{fig:soliton_molecule}
     }
\end{figure}

{\it Chiral Solitonic Molecules---} 
Even more interesting phenomena appear when we consider a finite magnetic field, and the Hamiltonian becomes $\hat H_{\rm sK}\rightarrow \hat H_{\rm sK}+h\sum_n\hat{S}_n^z$. 
A finite magnetic field $h$ gives raise to coupled equations of motion for $J_{n}^{x}$ and $J_{n}^{y}$, 
\begin{equation}    \label{Eq:semiSh}
\begin{split}
    \partial_t J_{n}^{x} =& (\Delta + J)J_{n-1}^{x}J_{n}^{z} +(\Delta - J)J_{n+1}^{x}J_{n}^{z}-h J_{n}^{y},\\
    \partial_t J_{n}^{y} =& (J - \Delta)J_{n-1}^{y}J_{n}^{z} -(\Delta + J)J_{n+1}^{y}J_{n}^{z}+h J_{n}^{x}.
\end{split}
\end{equation}
In its presence,  the chiral solitons include excitations in both $J^x_n$ and $J^y_n$, and form a bound molecule as shown in Fig.~\ref{fig:soliton_molecule}. 
The molecule consists of one $x$-soliton centered around $n_x$ and one $y$-soliton centered around $n_y$ both bound together. 
When $n\gg n_x$, the solitonic molecule is dominated by the excitation in $J_{n}^x$, i.e., with decreasing $n$ from $+\infty$, $J_{n}^{z}$ decreases from $1$ and $J_{n}^{x}$ increases similar to a single $x$-soliton previously discussed. 
When $n\ll n_y$, excitations in $J_{n}^{y}$ dominate.
A finite magnetic field couples $J_{n}^{x}$ and $J_{n}^{y}$, and both $J_{n}^{x}$ and $J_{n}^{y}$ become finite for a generic $n$. 
With increasing $h$, the solitonic molecule becomes more tightly bound, with the separation of $x$- and $y$-solitons decreasing and their overlap increasing, but the same phenomenology persists (Supplementary Materials~\cite{supplemental}). 

It is worth pointing out that the direction in which the solitonic molecule travels depends on the orientation of the molecule under fixed boundary conditions. 
If $n_y<n_x$, if we define a polarization of the solitonic molecule, $p=n_x-n_y$, $p>0$ in Fig.~\ref{fig:soliton_molecule}(a,c). 
This solitonic molecule travels to the right. 
Reversing the sign of $p$, i.e., placing the $x$-soliton to the left of the $y$-soliton, $n_x<n_y$, $p<0$ and the solitonic molecule travels to the left, as shown in Fig.~\ref{fig:soliton_molecule}(b,d). 
We thus see that the polarization $p$ can be used as a unique tool to control chiral transport. 
If we change the boundary condition, the solitonic molecule also reverses its travel direction, similar to a single soliton (Supplementary Materials~\cite{supplemental}).

{\it Summary and Outlook---}
We have shown that spin Kitaev models, which are realizable in various experimental platforms, provide us with intriguing phenomena unattainable in conventional spin models, including polarization-dependent chiral transport and chiral solitons. 
Our work holds promising applications for various quantum technologies. 
By tuning the ratio $\Delta/J$, one can manipulate the underlying curvature, enabling experimental simulations of quantum transport in curved geometries. 
The chiral nature of the solitons offers potential for developing nonreciprocal devices in quantum spintronics and quantum networks, where the solitons may serve as the foundation for robust quantum interferometry and transport, exhibiting resilience against external perturbations and disorder.

\nocite{Haake1987}

\begin{acknowledgments}
\noindent{
\textit{Acknowledgements---}
QZ acknowledges useful discussions with Lawrence Cheuk and Zoe Yan about experimental realizations of spin Kitaev model and XYZ model. We also thank David Wellnitz and Calder Miller for carefully review of the manuscripts and Rebekah Hermsmeier, and Timur Tscherbul for excellent discussions. QZ acknowledges supports by the National Science Foundation (NSF) through Grant No. PHY-2110614 and the Air Force Office of Scientific Research under award number FA9550-20-1-0221. AMR acknowledges  support from ARO grant W911NF24-1-0128, AFOSR MURI FA9550-21-1-0069, NSF JILA-PFC PHY-2317149 and NIST. 
}
\end{acknowledgments}

\onecolumngrid
\newpage
\vspace{0.4in}
\centerline{\bf\large Supplementary Materials for }
\centerline{\bf\large ``Polarization-dependent chiral transport and chiral solitons in spin Kitaev models"}
\setcounter{equation}{0}
\setcounter{figure}{0}
\setcounter{table}{0}
\makeatletter
\renewcommand{\theequation}{S\arabic{equation}}
\renewcommand{\thefigure}{S\arabic{figure}}
\renewcommand{\thetable}{S\arabic{table}}
\vspace{0.2in}

\section{Effective Hamiltonian of Floquet driving}

\begin{figure}
    \centering
    \includegraphics[width=0.75\linewidth]{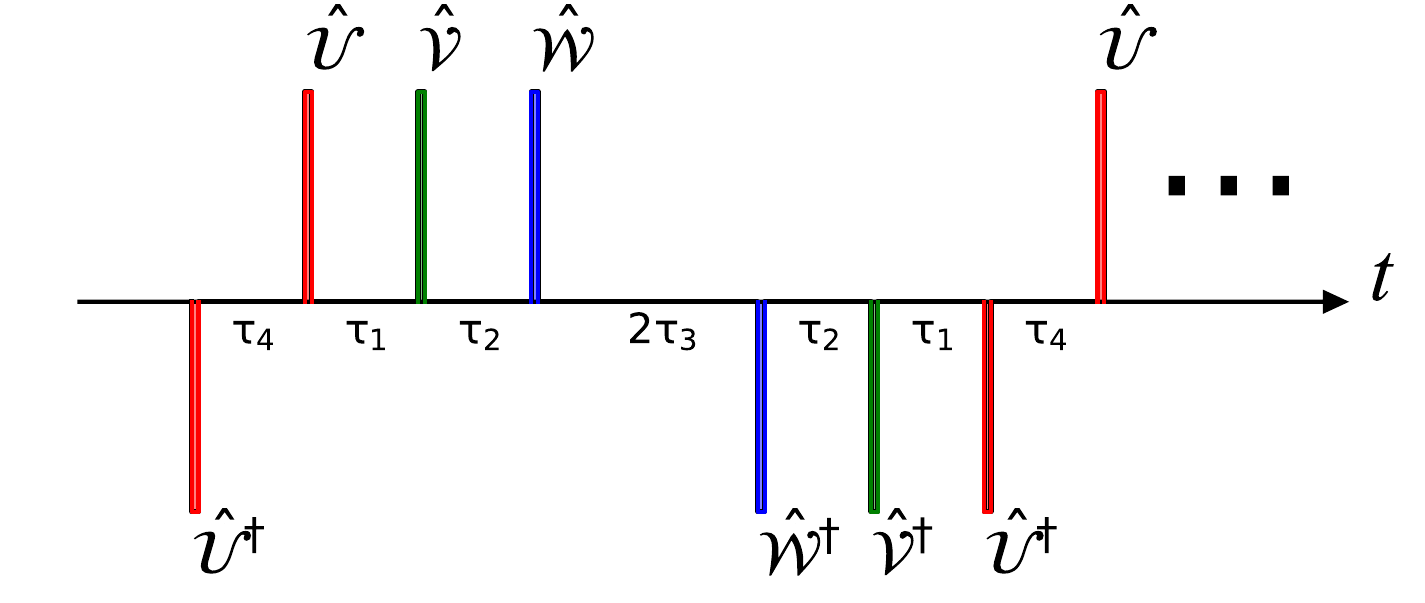}
    \caption{Floquet driving sequence of generating an effective Hamiltonian of the spin Kitaev model.}
    \label{figs:effective}
\end{figure}

This section discusses how to obtain the spin Kitaev model by Floquet driving. 
We consider the Hamiltonian $\hat{H}'_{\rm ex}=(J_0/2S)\sum_n(\hat{S}^+_n\hat{S}^-_{n+1}+h.c.)=(J_0/S)\sum_n (\hat{S}^x_n\hat{S}^x_{n+1}+\hat{S}^y_n\hat{S}^y_{n+1})$ and unitary operations $\hat{\mathcal{U}} = e^{-i\sum_{m}m \hat S_m^z\pi/2}$, $\hat{\mathcal{V}}=\prod_{m\in 2\mathbb{Z}}e^{-i\hat S_m^x \pi}$ and $\hat{\mathcal{W}}=\prod_{m\in 2\mathbb{Z}}e^{-i\hat S_m^z \pi/2}$. 
$\hat{\mathcal{U}}$ corresponds to a position-dependent rotation about the $z$ axis. 
$\hat{\mathcal{V}}$ and $\hat{\mathcal{W}}$ are $\pi$ and $\pi/2$-pulses applied to even sites about the $x$ and $z$ axis, respectively. 
It is straightforward to obtain
\begin{equation}
    \hat{\mathcal{V}}^\dag \hat{H}'_{\rm ex} \hat{\mathcal{V}} = \frac{J_0}{S}\sum_n (\hat{S}^x_n\hat{S}^x_{n+1}-\hat{S}^y_n\hat{S}^y_{n+1}),
\end{equation}
\begin{equation}
    \hat{\mathcal{V}}^\dag\hat{\mathcal{W}}^\dag \hat{H}'_{\rm ex} \hat{\mathcal{W}}\hat{\mathcal{V}} = \frac{J_0}{S}\sum_n (\hat{S}^y_n\hat{S}^x_{n+1}+\hat{S}^x_n\hat{S}^y_{n+1}),
\end{equation}
and
\begin{equation}
    \hat{\mathcal{U}} \hat{H}'_{\rm ex} \hat{\mathcal{U}}^\dag = \frac{J_0}{S}\sum_n (\hat{S}^x_n\hat{S}^y_{n+1}-\hat{S}^y_n\hat{S}^x_{n+1}). 
\end{equation}
We consider the Floquet driving sequence shown in Fig.~\ref{figs:effective}. 
The propagator from $t=0$ to $t=T$ is written as 
$\hat U=e^{-i\hat H_{\rm eff}T}$ and 
\begin{equation}
\begin{split}
    \hat U=& \hat{\mathcal{U}} e^{-i\hat{H}'_{\rm ex}\tau_4}\hat{\mathcal{U}}^\dag e^{-i\hat{H}'_{\rm ex}\tau_1}\hat{\mathcal{V}}^\dag e^{-i\hat{H}'_{\rm ex}\tau_2}\hat{\mathcal{W}}^\dag e^{-2i\hat{H}'_{\rm ex}\tau_3}\hat{\mathcal{W}} e^{-i\hat{H}'_{\rm ex}\tau_2}\hat{\mathcal{V}} e^{-i\hat{H}'_{\rm ex}\tau_1}\hat{\mathcal{U}}e^{-i\hat{H}'_{\rm ex}\tau_4}\hat{\mathcal{U}}^\dag\\ 
    =& (\hat{\mathcal{U}} e^{-i\hat{H}'_{\rm ex}\tau_4}\hat{\mathcal{U}}^\dag) e^{-i\hat{H}'_{\rm ex}\tau_1}(\hat{\mathcal{V}}^\dag e^{-i\hat{H}'_{\rm ex}\tau_2}\hat{\mathcal{V}} )(\hat{\mathcal{V}}^\dag \hat{\mathcal{W}}^\dag e^{-2i\hat{H}'_{\rm ex}\tau_3}\hat{\mathcal{W}}\hat{\mathcal{V}}) (\hat{\mathcal{V}}^\dag e^{-i\hat{H}'_{\rm ex}\tau_2}\hat{\mathcal{V}}) e^{-i\hat{H}'_{\rm ex}\tau_1}(\hat{\mathcal{U}}e^{-i\hat{H}'_{\rm ex}\tau_4}\hat{\mathcal{U}}^\dag).\\ 
\end{split}
\end{equation}
The effective Hamiltonian is written as 
\begin{equation}
    \hat H_{\rm eff}= \sum_n \frac{2J_0}{ST}\begin{pmatrix}  \hat S_n^x& \hat S_n^y\end{pmatrix}
      \begin{pmatrix}
        \tau_1+\tau_2 & \tau_3+\tau_4 \\
        \tau_3-\tau_4 & \tau_1-\tau_2
    \end{pmatrix}
      \begin{pmatrix}  \hat S_{n+1}^x \\ \hat S_{n+1}^y\end{pmatrix},
\end{equation}
where $T=2(\tau_1+\tau_2+\tau_3+\tau_4)$. 
Comparing it with the Hamiltonian of the spin Kitaev model, we find 
\begin{equation}
    \begin{split}
        \frac{2J_0}{T}(\tau_1+\tau_2) = & \Delta \cos(\phi) + J \cos(\theta),\\
        \frac{2J_0}{T}(\tau_1-\tau_2) = & -\Delta \cos(\phi) + J \cos(\theta),\\
        \frac{2J_0}{T}(\tau_3-\tau_4) = & \Delta \sin(\phi) - J \sin(\theta),\\
        \frac{2J_0}{T}(\tau_3+\tau_4) = & \Delta \sin(\phi) + J \sin(\theta).
    \end{split}
\end{equation}
Tuning $\tau_{i=1,2,3,4}$ thus allows experimentalists to change the amplitudes and phases of the flip-flop and flip-flip(flop-flop) terms. 

\section{Floquet driving scheme for realizing higher $S$ via Heisenberg interactions}

We propose to realize effective larger spin $S$ models starting from ensembles of spin $1/2$'s confined in dimensions transverse to the Kitaev chain of interest, e.g. a two-dimensional plane transverse to the one-dimensional Kitaev chain.

For spins $\hat{S}_{n,\alpha}$, where $n$ labels the position along the Kitaev chain, and $\alpha$ the additional transverse positions, we may define a collective spin $\hat{S}_n = \sum_{\alpha} \hat{S}_{n,\alpha}$ with size $S = N/2$ for $N$ spin 1/2's in the transverse dimension. 
Strong intra-plane Heisenberg interactions $V_{(n,\alpha),(n,\beta)} \, \hat S_{n, \alpha} \cdot \hat S_{n, \beta}$, where $V_{(n,\alpha),(n,\beta)}$ is the coupling between spins at the same chain position $n$, but different positions $\alpha, \beta$ in the transverse dimension, couple the spins in the transverse dimension. 
For sufficiently strong and long-range interactions in this transverse dimension, these Heisenberg interactions protect the collective Dicke manifold of the spin for initially fully polarized states, making the individual spins in a layer behave collectively and keeping the dynamics within the Dicke manifold~\cite{Rey2008S, Bilitewski2023S}. 
Projecting the interactions onto the collective manifold the intra-plane interactions do not affect the dynamics and may be dropped, whereas the interactions between different Kitaev chain positions $(n,n+1)$ are given by the average over the transverse direction $V_{n,n+1} = \sum_{\alpha,\beta} V_{(n,\alpha),(n+1,\beta)}$.

To apply this procedure, we therefore require a Floquet scheme that engineers the desired inter-plane Kitaev model from exchange interactions, while realizing intra-plane fully symmetric Heisenberg interactions. 
We start from the exchange Hamiltonian separated into intra-plane and inter-plane interactions (for simplicity dropping the spatial dependence $V_{(n,\alpha),(m,\beta)}$) 
\begin{equation}
    \begin{split}
        \hat{H}_{\mathrm{ex}} 
        &= \sum_n \hat{H}_{\mathrm{ex,intra}}^{(n)} + \sum_n \hat{H}_{\mathrm{ex,inter}}^{(n)} \\
        &= \sum_n \sum_{\alpha \neq \beta }  \bigg(\hat{S}^x_{(n,\alpha)} \hat{S}^x_{(n,\beta)} + \hat{S}^y_{(n,\alpha)}\hat{S}^y_{(n,\beta)}\bigg) + \sum_n \sum_{\alpha , \beta} \bigg(\hat{S}^x_{(n,\alpha)} \hat{S}^x_{(n+1,\beta)} + \hat{S}^y_{(n,\alpha)}\hat{S}^y_{(n+1,\beta)}\bigg),
    \end{split}
\end{equation}
where the first sum is the intra-plane interactions coupling spins in the same layer $n$, and the second sum describes the exchange interactions between neighboring layers $n$ and $n+1$.

Defining the pulse-operators $\hat{U}_x = \prod_m e^{i \pi/2 \hat{S}^x_m}$ and $\hat{V}_y = \prod_m e^{i (-1)^m \pi/2 \hat{S}^y_m}$ we note
\begin{align}
\hat{U}_x^{\dagger} \hat{H}_{\mathrm{ex,intra}}^{(n)} \hat{U}_x &= \sum_{\alpha \neq \beta} \hat{S}^x_{(n,\alpha)} \hat{S}^x_{(n,\beta)} + \hat{S}^z_{(n,\alpha)}\hat{S}^z_{(n,\beta)},\\
\hat{V}_y^{\dagger} \hat{H}_{\mathrm{ex,intra}}^{(n)} \hat{V}_y &= \sum_{\alpha \neq \beta} \hat{S}^z_{(n,\alpha)} \hat{S}^z_{(n,\beta)} + \hat{S}^y_{(n,\alpha)}\hat{S}^y_{(n,\beta)},
\end{align}
and
\begin{align}
\hat{U}_x^{\dagger} \hat{H}_{\mathrm{ex,inter}}^{(n)} \hat{U}_x &= \sum_{\alpha \beta} \hat{S}^x_{(n,\alpha)} \hat{S}^x_{(n+1,\beta)} + \hat{S}^z_{(n,\alpha)}\hat{S}^z_{(n+1,\beta)},\\
\hat{V}_y^{\dagger} \hat{H}_{\mathrm{ex,inter}}^{(n)} \hat{V}_y &= \sum_{\alpha \beta}   \hat{S}^y_{(n,\alpha)}\hat{S}^y_{(n+1,\beta)}-\hat{S}^z_{(n,\alpha)} \hat{S}^z_{(n+1,\beta)},
\end{align}
and thus obtain the effective Hamiltonian
\begin{equation}
    \begin{split}
        H_{\mathrm{eff}} 
         &= (\hat{H}_{\mathrm{ex}} + \hat{U}_x^{\dagger} \hat{H}_{\mathrm{ex}} \hat{U}_x + \hat{V}_y^{\dagger} \hat{H}_{\mathrm{ex}} \hat{V}_y )/3\\
         &= \frac{2}{3} \sum_n \sum_{\alpha \neq \beta }  \bigg(\hat{S}^x_{(n,\alpha)} \hat{S}^x_{(n,\beta)} + \hat{S}^y_{(n,\alpha)}\hat{S}^y_{(n,\beta)} + \hat{S}^z_{(n,\alpha)}\hat{S}^z_{(n,\beta)} \bigg) + \frac{2}{3} \sum_n \sum_{\alpha , \beta}  \bigg(\hat{S}^x_{(n,\alpha)} \hat{S}^x_{(n+1,\beta)} + \hat{S}^y_{(n,\alpha)}\hat{S}^y_{(n+1,\beta)}\bigg).
    \end{split}
\end{equation}
We may then perform the pulse sequence described in the previous section to further Floquet engineer the inter-plane interactions to realize the desired spin Kitaev Hamiltonian. 
Since the pulses described there act as global spin rotations within a given layer $n$, they commute with and do not affect the fully rotationally symmetric in-layer Heisenberg interactions engineered here.

\section{Polarization-dependent spin dynamics for different $S$}

\begin{figure}
    \centering
    \includegraphics[width=0.75\linewidth]{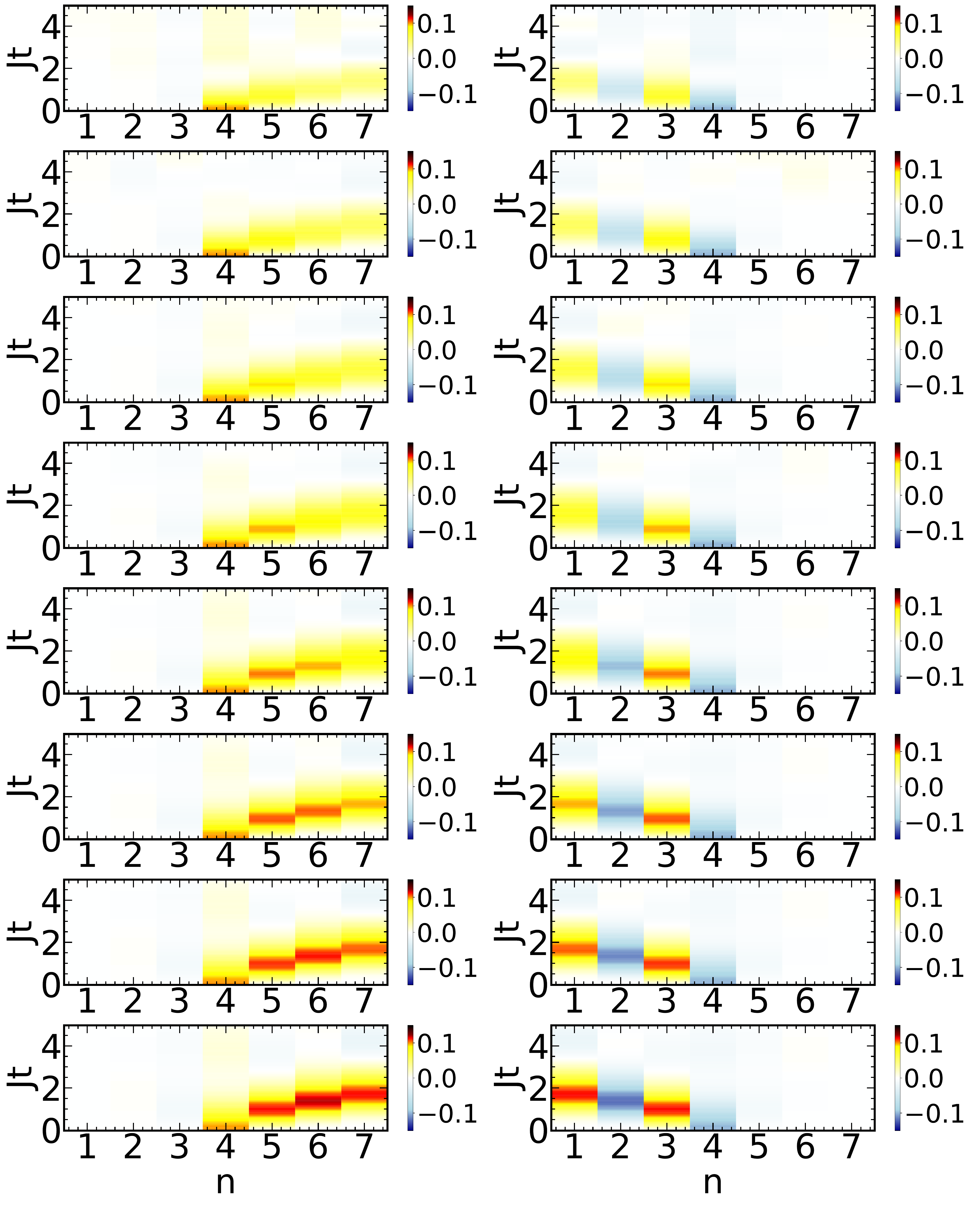}
    \caption{Polarization-dependent chiral spin transport. 
    $J^x_n(t)$ (left panels) and  $J^y_n(t)$ (right panels) as functions of time for spin $S=1, 3/2,\dots,4,9/2$ from top to bottom. 
    $\Delta/J=0.8$ and $N=7$ have been used. }
    \label{figs:dynamics}
\end{figure}

In Fig.~\ref{figs:dynamics}, we show the numerical results of polarization-dependent spin dynamics for all total spins between $1/2$ and $5$ in a lattice of $N=7$ sites.
As we did in the main text, $n_0=4$ and the initial excitations in $S^x_{n_0}$ and  $S^y_{n_0}$ are created by $e^{-i\alpha \hat{S}^y_{n_0}}$ or $e^{-i\alpha \hat{S}^x_{n_0}}$ with $\alpha = 0.1$, respectively.
The asymmetric propagation is clear for all these values of $S$. 
The amplification becomes evident as $S$ increases, which will become the exponential enhancement in the Hatano-Nelson model when $S\to\infty$.

\section{Mean field approach}

The Heisenberg equations of motion are written as 
\begin{equation}
    \begin{split}
       \partial_t{\hat S^x}_n=i[\hat H_{\rm sK},\hat S^x_n]=&-h \hat S^y_n+\frac1{S}[J\cos(\theta)-\Delta\cos(\phi)](\hat S^y_{n-1}+\hat S^y_{n+1})\hat S^z_n\\
        &+\frac{1}{S}J\sin(\theta)(\hat S^x_{n-1}-\hat S^x_{n+1})\hat S^z_n+\frac{1}{S}\Delta\sin(\phi)(\hat S^x_{n-1}+\hat S^x_{n+1})\hat S^z_n,\\
        \partial_t{\hat S^y}_n=i[\hat H_{\rm sK},\hat S^y_n]=&+h \hat S^x_n-\frac1{S}[J\cos(\theta)+\Delta\cos(\phi)](\hat S^x_{n-1}+\hat S^x_{n+1})\hat S^z_n\\
        &+\frac{1}{S}J\sin(\theta)(\hat S^y_{n-1}-\hat S^y_{n+1})\hat S^z_n-\frac{1}{S}\Delta\sin(\phi)(\hat S^y_{n-1}+\hat S^y_{n+1})\hat S^z_n,\\
        \partial_t{ \hat S^z}_n=i[\hat H_{\rm sK},\hat S^z_n]=&
        i\frac1{2S}\bigg(Je^{-i\theta}\hat S^+_{n+1}\hat S^-_n-Je^{i\theta}\hat S^-_{n+1}\hat S^+_n
        -\Delta e^{-i\phi}\hat S^+_{n+1}\hat S^+_n+\Delta e^{i\phi}\hat S^-_{n+1}\hat S^-_n\bigg)\\
        &+i\frac1{2S}\bigg(-Je^{-i\theta}\hat S^+_{n}\hat S^-_{n-1}+Je^{i\theta}\hat S^-_{n}\hat S^+_{n-1}
        -\Delta e^{-i\phi}\hat S^+_{n}\hat S^+_{n-1}+\Delta e^{i\phi}\hat S^-_{n}\hat S^-_{n-1}\bigg).
    \end{split}
\end{equation}
We define $\langle \hat S^\mu_{n}\rangle /S=J^\mu_{n}$. In the large $S$-limit, the operators are replaced by their expectation values 
as quantum fluctuation $\sim1/S$ ~\cite{Haake1987S}. 
We thus obtain the equations of motion for the magnetization $J_n^\mu$,
\begin{equation}
    \begin{split}
       \dot J^x_n(t)=&-hJ^y_n+[J\cos(\theta)-\Delta\cos(\phi)](J^y_{n-1}+J^y_{n+1})J^z_n+J\sin(\theta)(J^x_{n-1}-J^x_{n+1})J^z_n+\Delta\sin(\phi)(J^x_{n-1}+J^x_{n+1})J^z_n,\\
        \dot J^y_n(t)=&+hJ^x_n-[J\cos(\theta)+\Delta\cos(\phi)](J^x_{n-1}+J^x_{n+1})J^z_n+J\sin(\theta)(J^y_{n-1}-J^y_{n+1})J^z_n-\Delta\sin(\phi)(J^y_{n-1}+J^y_{n+1})J^z_n,\\
        \dot J^z_n(t)
         =& 
         (J_{n-1}^x J_{n}^y+J_{n+1}^x J_{n}^y) (\Delta  \cos (\phi )+J \cos (\theta ))+J_{n-1}^x J_{n}^x (-\Delta  \sin (\phi )-J \sin (\theta ))+J_{n-1}^y J_{n}^y (\Delta  \sin (\phi )-J \sin (\theta ))\\
         +&(J_{n-1}^y J_{n}^x+J_{n+1}^y J_{n}^x) (\Delta  \cos (\phi )-J \cos (\theta ))+J_{n+1}^x J_{n}^x (-\Delta  \sin (\phi )+J \sin (\theta ))+J_{n+1}^y J_{n}^y (\Delta  \sin (\phi )+J \sin (\theta )).
    \end{split}\label{Eqs:semi_full}
\end{equation}
When $\phi=\theta=\pi/2$, we obtain
\begin{equation}
    \begin{split}
        \dot J^x_n(t)=&-h J^y_n+(J+\Delta)J^x_{n-1}J^z_n+(\Delta-J) J^x_{n+1}J^z_n,\\
        \dot J^y_n(t)=&+h J^x_n+(J-\Delta)J^y_{n-1}J^z_n-(\Delta+J) J^y_{n+1}J^z_n,\\
        \dot J^z_n(t)=&
            J(J^{x}_{n+1}J^{x}_{n}+J^{y}_{n+1}J^{y}_{n})
            -\Delta(J^{x}_{n+1}J^{x}_{n}-J^{y}_{n+1}J^{y}_{n})
            -J(J^{x}_{n-1}J^{x}_{n}+J^{y}_{n-1}J^{y}_{n})
            -\Delta(J^{x}_{n-1}J^{x}_{n}-J^{y}_{n-1}J^{y}_{n}).\label{sK:semi}
    \end{split}
\end{equation}
Since $(J_n^{x})^2+(J_n^{y})^2+(J_n^{z })^2=1$, $J^{z}_n$ is replaced by $\pm\sqrt{1-(J^{x}_n)^2-(J^{y}_n)^2}$ and we obtain the equations of motion in the main text.

\section{Results of $\theta\neq \pi/2$}

When $\theta\neq \pi/2$, the chiral transport can still be understood from the interferences between multiple pathways at short times. 
To show this, we consider $ \ket{\psi(0)}=e^{-i\hat S_z \Phi_0}e^{-i\hat S_y\alpha}|F\rangle$ as the initial state. 
Similar to the main text, we consider three sites, $n_0-1$, $n_0$, and $n_0+1$. 
For a small $\alpha$, the initial state can be approximately written as 
\begin{equation}
    \ket{\psi(0)} = e^{-iS\Phi_0}\ket{...}\otimes(\ket{S,S,S}+c_1\ket{S,S-1,S}),
\end{equation}
where $c_1=\alpha e^{i\Phi_0}\sqrt{S/2}$, $\ket{S, S-1, S}$ is a short-hand notation for spin states in sites $n_0$ and $n_0\pm1$, $\ket{S, S-1, S}\equiv\ket{S}_{n_0-1}\ket{S-1}_{n_0}\ket{S}_{n_0+1}$, $\ket{...}\equiv\prod_{m\neq n_0,n_0\pm 1}|S\rangle_m$. 
At small $t$, $e^{-i\hat H_{\rm sK}t}\approx 1-i (\hat H_{\rm ex}+\hat H_{\rm ff})t$. Two pathways of spin transport thus exist, one induced by $\hat H_{\rm ex}$ and the other by $\hat H_{\rm ff}$. 
It is clear that 
\begin{equation}
    \hat H_{\rm ex} \ket{S,S-1,S} \sim e^{-i\theta} \ket{S-1,S,S}+e^{i\theta} \ket{S,S,S-1},
\end{equation}
and 
\begin{equation}
    \hat H_{\rm ff} \ket{S,S,S} \sim e^{i\phi}\ket{S-1,S-1,S}+ e^{i\phi} \ket{S,S-1,S-1}.
\end{equation}
The final state reads 
\begin{equation}
    \ket{\psi(t)} = \ket{...}\otimes\left(\psi_0+\psi_1(t)+\psi_2(t)\right)~\label{Eqs:interference}
\end{equation}
where 
\begin{equation}
\begin{split}
&\psi_0 = \ket{S,S,S}+c_1\ket{S,S-1,S},\\
&\psi_1(t)= -i Jt c_1( e^{i\theta}\ket{S,S,S-1} + e^{-i\theta}\ket{S-1,S,S} ),\\
&\psi_2(t)=-i \Delta t(e^{i\phi}\ket{S-1,S-1,S}+e^{i\phi}\ket{S,S-1,S-1}).
\end{split}
\end{equation}
Putting together the contributions from $\psi_1(t)$ and $\psi_2(t)$, we obtain
\begin{equation}
    \begin{split}
        S_{n_0-1}^x = & \alpha St [J\sin(-\theta+\Phi_0) + \Delta\sin(\phi-\Phi_0)],\\
        S_{n_0+1}^x = & \alpha St [J\sin(+\theta+\Phi_0) + \Delta\sin(\phi-\Phi_0)],\\
    \end{split}
\end{equation}
and
\begin{equation}
    \begin{split}
        S_{n_0-1}^y = & -\alpha St [J\cos(-\theta+\Phi_0) + \Delta\cos(\phi-\Phi_0)],\\
        S_{n_0+1}^y = & -\alpha St [J\cos(+\theta+\Phi_0) + \Delta\cos(\phi-\Phi_0)].\\
    \end{split}
\end{equation}
When $\phi=\theta=\pi/2$, $\Phi_0=0$ or $\pi/2$, the expressions above reduce to the results in the main text.
Polarization-dependent constructive/destructive interference always exists when both $J$ and $\Delta\neq 0$ for a finite $\theta$.

As for the solitonic solution, we consider a linear combination of $J^{x,y}_n$,
\begin{equation}
    J^\pm_n = \cos(\Phi^\pm)J^x_n + \sin(\Phi^\pm)J^y_n,
\end{equation}
where 
\begin{equation}
    \Phi^\pm = \frac{\phi}{2}\pm \frac{1}{2}\arccos(\frac{J\cos(\theta)}{\Delta}),
\end{equation}
such that the semi-classical equations of motion Eqs.~(\ref{Eqs:semi_full}) at $h=0$ could be decoupled,
\begin{equation}
    \dot J^\pm_n = (t_R^\pm J^\pm_{n-1} + t_L^\pm J^\pm_{n+1}) J^z_n,\label{Eqs:decoupled_semi}
\end{equation}
with
\begin{equation}
    t_R^{\pm}=+J \sin (\theta )\mp\sqrt{\Delta ^2-J^2 \cos ^2(\theta )},\quad 
    t_L^{\pm}=-J \sin (\theta )\mp\sqrt{\Delta ^2-J^2 \cos ^2(\theta )}.
\end{equation}
Our discussions in the main text apply and $J^\pm_n$ shows polarization-dependent chiral transport when $\Delta>|J\cos(\theta)|$ since $|t_R^{+}| > |t_L^{+}|$ and $|t_R^{-}| < |t_L^{-}|$.
Comparing with $\theta=\phi=\pi/2$, we find $J$ and $\Delta$ are rescaled to $J\sin(\theta)$ and $\sqrt{\Delta^2-J^2\cos^2(\theta)}$, respectively.

To obtain the spatial profiles of the solitons, we numerically solve Eq.~(\ref{Eqs:semi_full}) by choosing $J^\mu_n(t)=J^{\mu}_{n-v_st_0}(t-t_0)=\mathcal{J}^\mu(s = n-v_st)$, where $\mathcal{J}^\mu(s)$ slowly varies over $n$.
Solving Eq.~(\ref{Eqs:semi_full}), therefore, amounts to integrating
\begin{equation}
   \begin{split}
    \partial_s\mathcal{J}^x =& -\frac{-h \mathcal{J}^y + 2 {\mathcal{J}^z} ({\mathcal{J}^y} (J \cos (\theta )-\Delta  \cos (\phi ))+\Delta  {\mathcal{J}^x} \sin (\phi ))}{v_s-2 J \sin (\theta ) {\mathcal{J}^z}},\\
    \partial_s\mathcal{J}^y =& +\frac{-h \mathcal{J}^x + 2 {\mathcal{J}^z} ({\mathcal{J}^x} (\Delta  \cos (\phi )+J \cos (\theta ))+\Delta  {\mathcal{J}^y} \sin (\phi ))}{v_s-2 J \sin (\theta ) {\mathcal{J}^z}},\\
    \partial_s\mathcal{J}^z =& \frac{2 \Delta  \left(\sin (\phi ) \left(({\mathcal{J}^x})^2-({\mathcal{J}^y})^2\right)-2 {\mathcal{J}^x} {\mathcal{J}^y} \cos (\phi )\right)}{v_s-2 J \sin (\theta ) {\mathcal{J}^z}},
   \end{split}
\end{equation}
with suitable boundary conditions at $s=\pm \infty$.
Equivalently, we use the spherical coordinates
\begin{equation}
    \mathcal{J}^x(s) = \sin(\Theta(s))\cos(\Phi(s)),\quad 
    \mathcal{J}^y(s) = \sin(\Theta(s))\sin(\Phi(s)),\quad 
    \mathcal{J}^z(s) = \cos(\Theta(s)),
\end{equation} 
such that the equations
\begin{equation}
    \begin{split}
        \partial_s\Theta =& -\frac{2 \Delta  \sin (\Theta) \sin (\phi -2 \Phi)}{v_s-2 J \sin (\theta ) \cos (\Theta)},\\
    \partial_s\Phi =& \frac{-h+2 \cos (\Theta) (J \cos (\theta )+\Delta  \cos (\phi -2 \Phi))}{v_s-2 J \sin (\theta ) \cos (\Theta)},
    \end{split}\label{eqs:soliton_spherical}
\end{equation}
are integrated in obtaining the initial soliton profiles.

\section{Boundary conditions of the solitons}

\begin{figure}[hb!]
    \centering
    \includegraphics[width=0.95\linewidth]{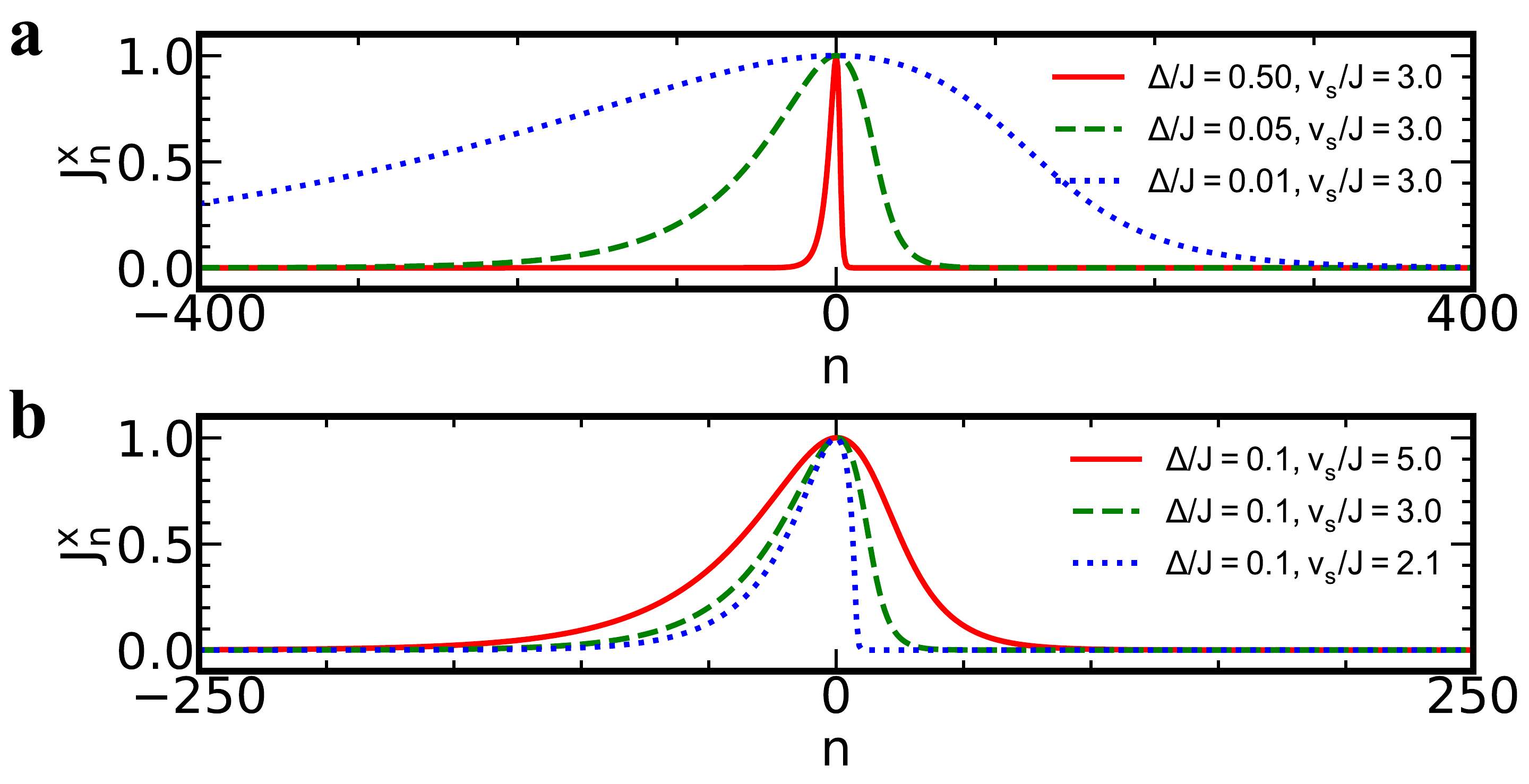}
    \caption{The dependence of profiles of $x$-solitons on  $\Delta$ and $v_s$. (a) $v_s$ is fixed at $2.5J$ and $\Delta$ decreases.  $\Delta/J = 0.5, 0.1, 0.05$ are represented by red solid, green dashed, and blue dotted curves. (b) $\Delta$ is fixed at $0.1J$ and $v_s$ decreases.  $v_s/J = 5, 3, 2.1$ are represented by red solid, green dashed, and blue dotted curves.}
    \label{figs:profile}
\end{figure}

\begin{figure}[hb!]
    \centering
    \includegraphics[width=0.75\linewidth]{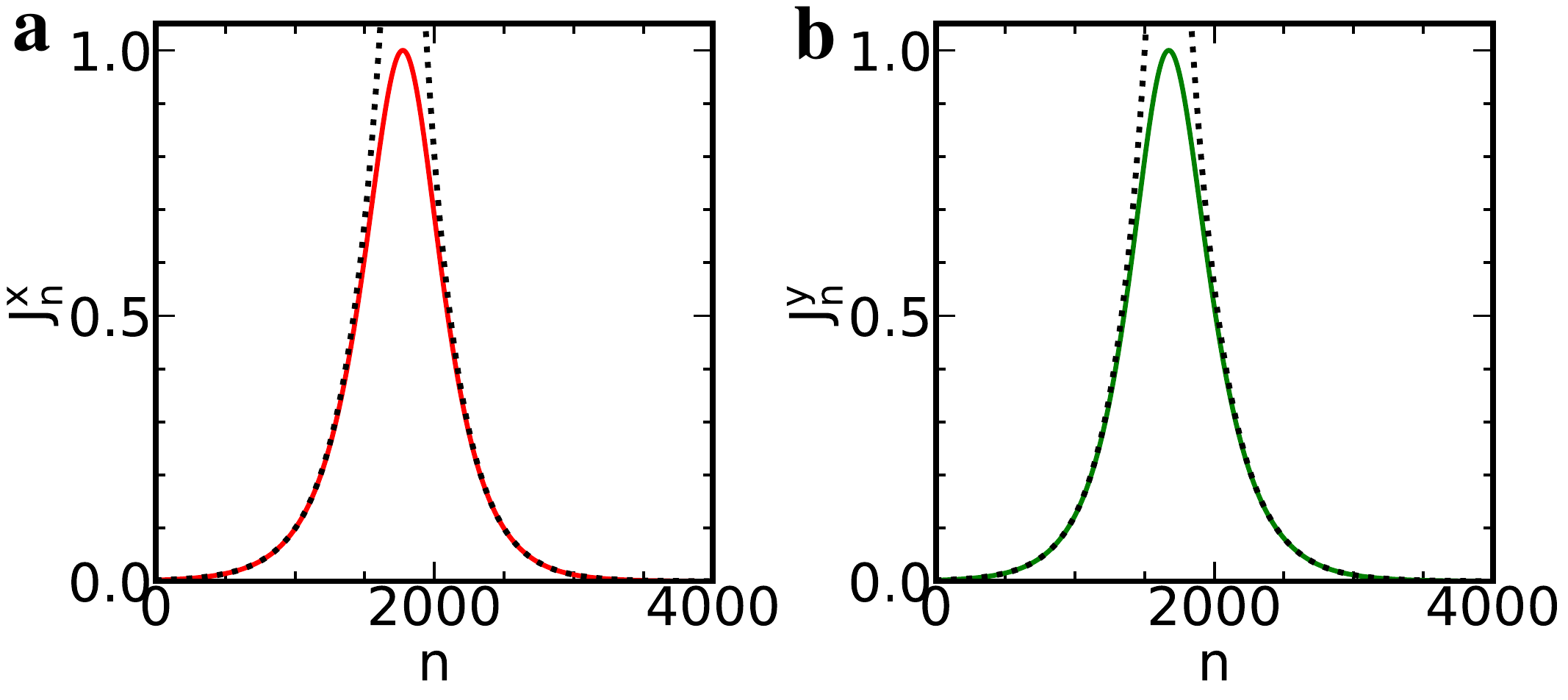}
    \caption{
    Comparisons between profiles of solitons (solid curves) and the solution to the Hatano-Nelson model (dashed curves). 
    (a) The $x$-soliton propagating towards right with $v_s=+50 J$. 
    (b) The $y$-soliton propagating towards left with $v_s=-50 J$. 
    The boundary condition is $J^z(\pm\infty)=\pm 1$.  $\Delta = 0.1J$, and $h=0$. }
    \label{figs:edge}
\end{figure}

In this section, we consider the edges of the solitons where the excitation amplitudes are small.
When $n\rightarrow -\infty$, $J_n^z\approx -1$, the equation of motion for $J_n^x$ in Eq.~(7) in the main text becomes
\begin{equation}
    \partial_t J_{n}^{x} = -[(\Delta + J)J_{n-1}^{x} +(\Delta - J)J_{n+1}^{x}],
    \label{Eq:semi_small_neg}
\end{equation}
where $\Delta>0, J>0$. The solution to Eqs.~\ref{Eq:semi_small_neg} is written as 
$J^x_n=e^{k^x_{-}(n-v_st)}$, and 
\begin{equation}
        -v_s k^x_- =-[(\Delta+J)e^{-k^x_-}+(\Delta-J)e^{+k_-^x}].
\end{equation}
Whereas $k^x_{-}$ can be obtained from the above equation numerically for any $v_s$, it has a simple analytic solution when the profile of the soliton changes slowly in the real space, i.e., when $|k^x_{-}|\ll 1$, 
\begin{equation}
    k^x_- = \frac{2\Delta}{2J + v_s}.
\end{equation}
$k^x_{-}>0$ need to be satisfied such that $J^{x}_n$ arises when $n$ increases from $-\infty$. 
When $n\rightarrow \infty$, $J_n^z\approx 1$, the equation of motion for $J_n^x$ is written as 
\begin{equation}
    \partial_t J_{n}^{x} = +[(\Delta + J)J_{n-1}^{x} +(\Delta - J)J_{n+1}^{x}].
    \label{Eq:semi_small_pos}
\end{equation}
The solution to Eqs.~\ref{Eq:semi_small_pos} is $J^x_n=e^{k^x_{+}(n-v_st)}$ and 
\begin{equation}
    k^x_+ = \frac{2\Delta}{2J - v_s},
\end{equation}
when $|k^x_{+}|\ll 1$. 
$k^x_+<0$ such that $J^{x}_n$ eventually decreases down to zero when $n\rightarrow \infty$.
These conditions above immediately yield $v_s>2J$ and for the $x$-solitons. 
Namely, an $x$-soliton moves towards the right when the boundary condition is fixed as $J_{n\to\pm\infty}^z\approx \pm 1$.
On the one hand, $k^x_\pm$ decreases with decreasing $\Delta$, leading to a broadening of the soliton profile (Fig.~\ref{figs:profile}(a-d)). 
In the $\Delta\to 0$ limit, the width of the soliton becomes infinite and the spatial profile becomes uniform. The solitonic solutions thus vanish.  
On the other hand, $k^x_+$ increases with decreasing $v_s$. 
As the soliton velocity $v_s\to 2J$,  $|k^x_+|$ diverges, and the wavefront of the soliton becomes infinitely sharp (Figs.~\ref{figs:profile}(e-h)). 
When $v_s \leq 2J$, it is impossible to satisfy the boundary condition, which requires that the profile decreases down to zero when $n\rightarrow \infty$. 
The soliton solutions thus cease to exist. 

Under the same boundary condition, we find the $y$-solitons satisfy 
\begin{equation}
    k^y_-=\frac{2\Delta}{-2J - v_s}>0,\quad k^y_+ = \frac{2\Delta}{-2J + v_s}<0.
\end{equation}
These conditions show $v_s<-2J$ for the $y$-solitons. 
A $y$-soliton moves towards the left.
In Fig.~\ref{figs:edge}, we compare the soliton solutions and the exact solutions for small excitations near the edges, and they show good agreement. 

\begin{figure}[h]
    \includegraphics[width=0.75\linewidth]{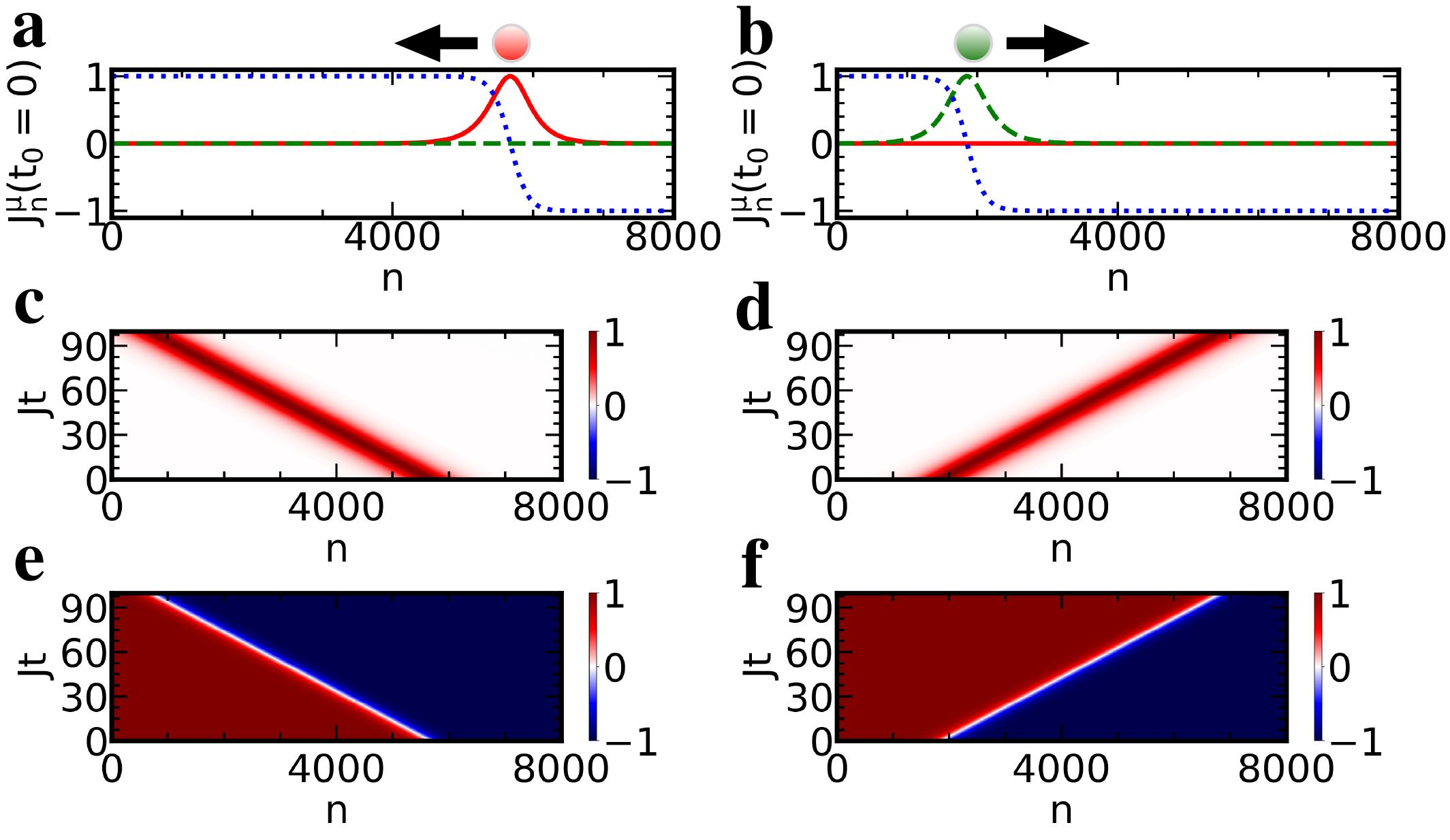}
    \caption{
    (a,b) Red solid, green dashed, and blue dotted curves are $J^x_n(t)$, $J^y_n(t)$, and $J^z_n(t)$ at $t_0=0$ of the (a) $x$-soliton (b) $y$-soliton, respectively.
    (c,e) Density plots of $J^x_n(t)$ and $J^z_n(t)$ of an $x$-soliton propagating towards left with $v_s=-50J$, $\Delta = 0.1J$, and $h=0$. 
    (d,f) Density plots of $J^y_n(t)$ and $J^z_n(t)$ of a $y$-soliton propagating towards right with $v_s=+50J$, $\Delta = 0.1J$, and $h=0$. }
    \label{figs:soliton}
\end{figure}

When the boundary condition changes into $J_{n\to\pm\infty}^z\approx \mp 1$, the conditions for $x$ and $y$-solitons change into $k^x_{+}>0$, $k^x_{-}<0$ and $k^y_{+}>0$, $k^y_{-}<0$. 
From the analysis above, we find that the conditions satisfied by the soliton velocity become $v_s<-2J$ and $v_s>2J$ for the $x$ and $y$-solitons. 
The corresponding soliton solutions are shown in Fig.~\ref{figs:soliton}.

\section{Boundary conditions and larger $h$}

\begin{figure}[h]
    \includegraphics[width=0.75\linewidth]{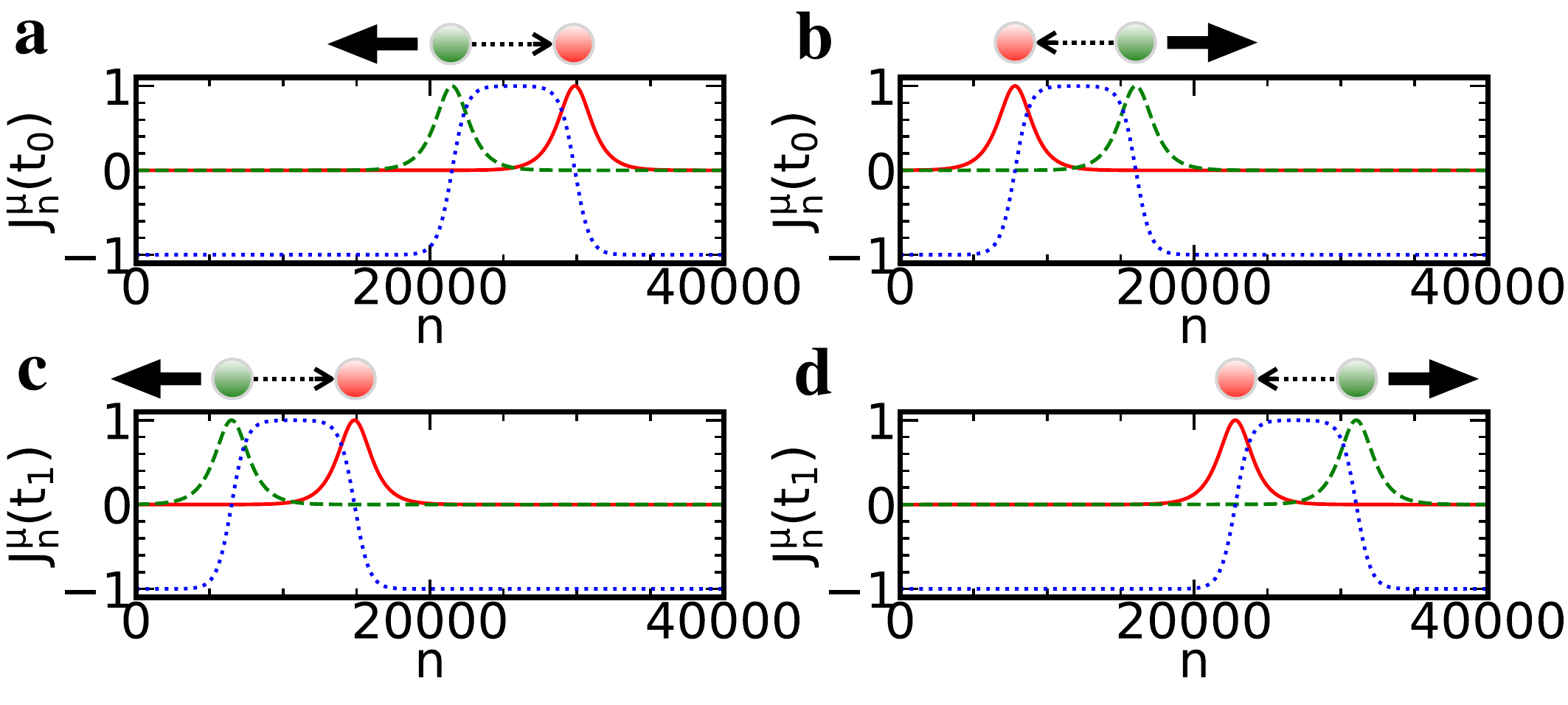}
    \caption{
    A solitonic molecule propagating towards the right (a,c) and the left (b,d) with $v_s=\pm 200J$. $\Delta = 0.1J$, and $h=+0.0001J$ are used. 
    The boundary condition is set as $J^z_{\pm\infty}=-1$. 
    Solid red, green dashed and blue dotted curves represent $J^x_n(t)$, $J^y_n(t)$ and $J^z_n(t)$ when $Jt_0=0$ (a,b) and $Jt_1 = 75$ (c,d). 
    Small dotted arrows denote the polarizations of solitonic molecules. 
    }
    \label{figs:soliton_molecule}
\end{figure}

\begin{figure}[h]
    \includegraphics[width=0.5\linewidth]{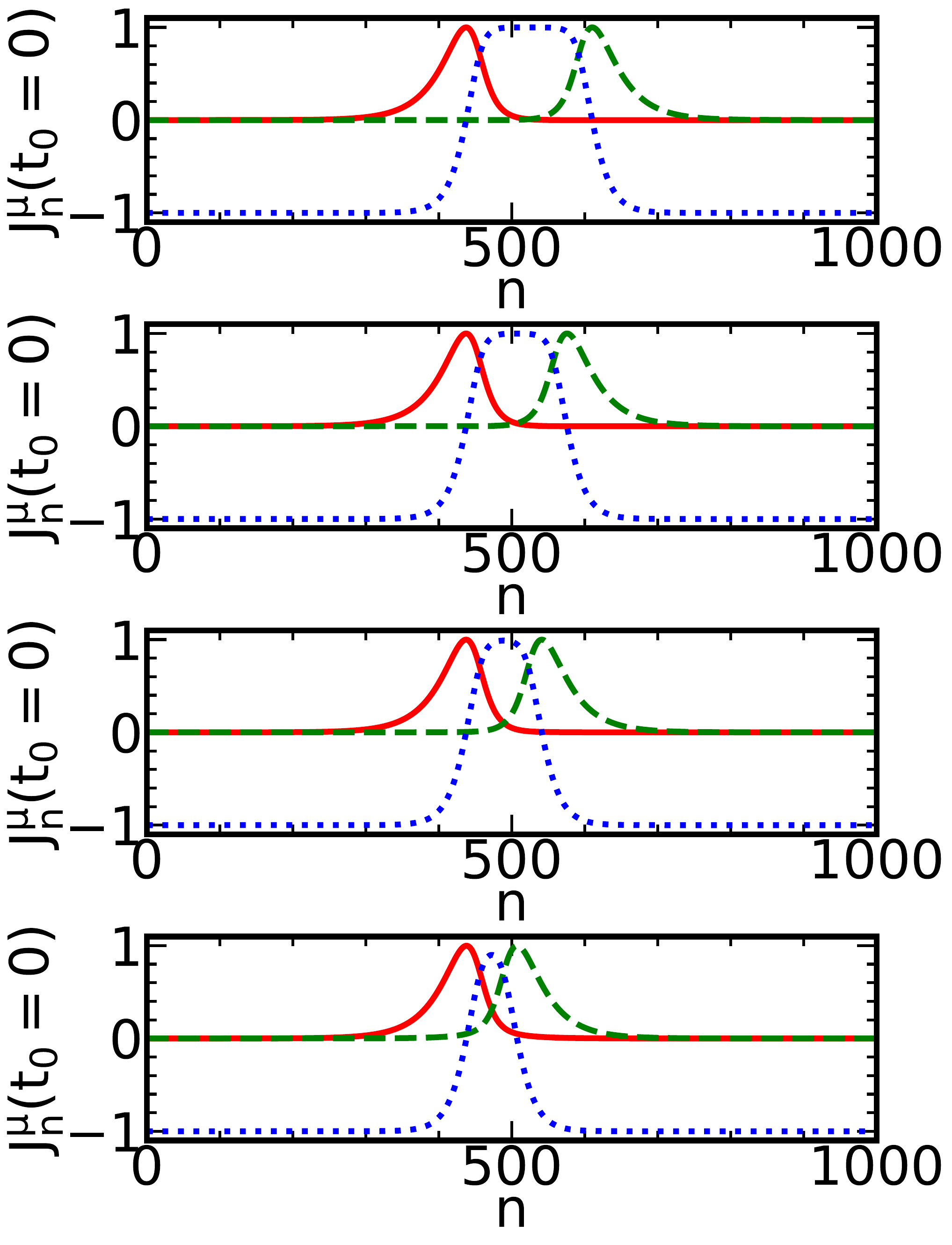}
    \caption{
    A solitonic molecule propagating towards the right with $v_s= 5J$. $\Delta = 0.1J$. The boundary condition is set as $J^z_{\pm\infty}=-1$. 
    Solid red, green dashed and blue dotted curves represent $J^x_n(t)$, $J^y_n(t)$ and $J^z_n(t)$ at $t_0=0$.
    From top to bottom: $h/J=10^{-5},10^{-4},10^{-3},10^{-2}$ are used.
    }
    \label{figs:molecule_h}
\end{figure}

Similar to the decoupled $x$- and $y$-solitons when $h=0$, the solitonic molecules change the directions of their velocities when the boundary condition is changed. 
In Fig.~\ref{figs:soliton_molecule}, we show the solitonic molecules with boundary conditions $J^z_{n\to\pm\infty}=-1$. 
The polarization of the solitonic molecule now points toward the opposite directions of their velocities. 
With increasing $h$, the $x$-soliton and the $y$-soliton become closer to each other and have stronger overlaps in a molecule, as shown in Fig.~\ref{figs:molecule_h}.  

\end{document}